\documentclass[aps,amsfonts,nofootinbib,superscriptaddress,preprint]{revtex4-1}
\usepackage{comment}
\usepackage[normalem]{ulem} %% Para tachar, comando sout \sout{for the first kinematics}
\makeatletter
\usepackage{amsmath}
\usepackage{amssymb}
\usepackage{appendix}
\usepackage{amsbsy}
\usepackage{bm}
\usepackage{epsfig} 
\usepackage{color}
\usepackage{slashed}
\usepackage{soul}
\usepackage{comment}
\makeatletter
\usepackage{amsmath}
\usepackage{amssymb}
\usepackage{appendix}
\usepackage{amsbsy}
\usepackage{bm}
\usepackage{epsfig} 
\usepackage{color}
\usepackage{slashed}
\newcommand{\stkout}[1]{\ifmmode\text{\sout{\ensuremath{#1}}}\else\sout{#1}\fi}

\newcommand{\ee}{\end{equation}}
\newcommand{\bb}{\begin{equation}}
\newcommand{\eqb}{\begin{eqnarray}}

\newcommand{\eqf}{\end{eqnarray}}

\def\pivec{\mbox{\boldmath$\pi$}}
\def\sigmavec{\mbox{\boldmath$\sigma$}}

\usepackage[T1]{fontenc}

\begin{document}
\title{  Supersymmetric Quantum Mechanics of  Continuous Topological Insulators}
\author{J. Gamboa}
\email{jorge.gamboa@usach.cl}
\affiliation{Departamento de F\'isica, Universidad de Santiago de Chile, Casilla 307, Santiago, Chile}
\author{F. M\'endez}
\email{fernando.mendez@usach.cl}
\affiliation{Departamento de F\'isica, Universidad de Santiago de Chile, Casilla 307, Santiago, Chile}

\begin{abstract}
We construct a class of quantum systems in a space continuum inspired by results from topological insulator physics. Instead of adding spin-orbit coupling terms suggested by time-reversal invariance as in conventional topological insulators, the terms ${\bf L}\cdot{\bf S}$ are determined using supersymmetry as a starting point. This procedure {{not only restricts the number of possible continuous topological insulators models but also provides a systematic way to find new continuous topological insulators models. Some explicit quantum mechanical examples are 
discussed and applications to dark matter physics are also outlined.}}
\end{abstract}
\maketitle

\section{ Introduction}

A topological insulator is a material that behaves like an insulator in bulk, but on the surface, electrons 
 can conduct electric current as in an ordinary conductor. On the surface of the insulator, the carriers are 
 Dirac fermions, but the conductivity depends on topological properties \cite{haldane}.
                                                                                                                                                                                                                                                                                                                                                                                                                                                                                                                                      
Notably, these systems show similar behavior to the FQHE but an external magnetic field is  unnecessary. The similarity with the FQHE is that the carriers also live in a two-dimensional edge similar to the FQHE with anyonic behavior if the sample is three-dimensional and as chiral bosons, if the insulator is two-dimensional \cite{wen} (see also \cite{moore,hansson}). 

From this point of view,  the analogy with the FQHE can be taken even further if we think of a topological insulator
 as a Hall hetero-junction where the magnetic field is not present, and the relevant question is, what is the analog of the 
 external magnetic field?

This problem was first studied in \cite{kane1},  who assumed that in the absence of an external magnetic field, the spin-orbit term plays this role (see also \cite{fu,moore1,roy}).

{{These authors reached this conclusion by noting that in analogy with the quantum Hall effect --which is invariant under time-reversal-- the only spin term that can be added to the Hamiltonian is the spin-orbit coupling one \cite{kane1,fu,moore1,roy}.

If we now assume that instead of a crystalline structure we have a continuous manifold, then we can define new classes of topological invariant insulators which could be used in applications of integrable models, high energy physics modeling or even be the starting point to study new physical phenomena.}}

%{{ In other words, instead of considering materials with well-defined crystal structures, we will focus on continuous fermionic systems that are invariant under time-reverse and show that they can be studied using techniques similar to those developed in the physics of topological insulators .

%In this paper we would like to propose that calculational techniques developed in topological insulator physics \cite{kane1,fu,moore1,roy} can also be used to solve other problems in high energy physics and mathematical physics. 

Particularly the spinning particles in external non-abelian gauge fields are the class of systems that can be analyzed from this perspective and their study having into account the subtleties of topological insulators physics might 
 shed light on problems that, as far as we know, have not been considered.

{{The theoretical construction of topological insulator models {{--as pointed out above -- requires time-reversal}} invariance, which provides the rule
\bb
H \to H_0 + \alpha {\bf L}\cdot{\bf S}, \label{rule}
\ee 
where $H_0 = -\frac{1}{2} \nabla^2 + u({\bf x})$, {{with $u$, the potential term}}.

However, although prescription (\ref{rule}) is correct, it leads to {{an infinitely large number of} topological insulators models and the question is how do we limit this arbitrariness.

There is an economically albeit physically strong possibility  to impose supersymmetry as a starting point, and this strategy is supported by a {{considerable amount}} of work \cite{01,lot,1,2,3,4}.

Then, let us start by considering the problem posed above with the help of techniques coming from supersymmetric quantum mechanics \cite{witten,review}. The advantages of {{ this procedure }} are at least three: the first is that the spin-orbit coupling {{comes}} from supersymmetrization,  and it is not a non-relativistic correction. Secondly, a  $SU(2)$ connection naturally emerges in the Hamiltonian, and third, the procedure proposed here might allow the analysis 
 of general systems whose properties are close to topological insulators models.
 
 This last aspect is interesting because it suggests that using supersymmetric quantum mechanics techniques  \cite{review} , we could solve a broad spectrum of  "supersymmetric topological insulators"  problems that can be useful in possible applications,  and in this paper, we would like to explain some examples in this direction.
 
 In three dimensions there are at least two different ways to supersymmetrize a physical system;  the first one is based on Clifford's 
 algebra for fermionic fields \cite{witten},  and the other,  on a modification that induces a spin-orbit coupling term for the
  Hamiltonian \cite{jj} (see also \cite{gozzi} for the one-dimensional case).  
  
  This last method of  supersymmetrization is the one we will use in this paper.  We start section II by explaining some essential aspects and other results that, as far as we know, are not discussed elsewhere.  In section III, we provide  two examples of how these ideas apply for the case of 
 a harmonic oscillator and the Aharonov-Bohm potentials; for the case of the oscillator in three dimensions, it turns
  out to be an extension of the system considered in \cite{zhang}, while for the second we show how we can build a topological insulator using the Aharonov-Bohm effect. {{In Section IV we provide two examples of continuous topological insulators connected with high energy physics, namely, an extension of Dirac's
 oscillator and the magnetoelectric effect in Dark Matter physics.}} The last section contains the conclusions.

\section{Supersymmetric Continuous Topological Insulators}

Let us summarize the main ideas of the supersymmetrization procedure  discussed in \cite{jj}.  For any 
quantum mechanical  system with Hamiltonian \footnote{In what follows, Hamiltonians are multiplied by the 
mass $m$, but we preserve the standard notation $H$.}
\bb
H = -\frac{1}{2} \nabla^2 + u({\bf x}), 
\label{01}
\ee
where $u({\bf x})$ is the potential, the ground state has no nodes and, therefore,  this particular state  can be written as 
\bb 
\psi_0=  e^{-V}, \label{f0}
\ee 
where $V({\bf x})>0$, and  real.

Then, the Schr\"odinger equation for the ground state,  $H \psi_0 = E_0 \psi_0$,  reads 
\bb 
{\bar H} \psi_0 =0, \label{f1}
\ee 
with ${\bar H} = H-E_0$.

Replacing (\ref{f0}) in (\ref{f1}), the Hamiltonian ${\bar H}$ can be written as follows 
\bb
{\bar H}= \frac12\,Q^\dagger_i Q_i,
\label{f2}
\ee
with $\{i,j\}\in\{1,2,3\}$ and  
\bb
Q^\dagger_i = -\partial_i + \partial_i V, ~~~Q_i= \partial_i + \partial_i V,
\label{f3}
\ee
and $ u=\frac12\left((\nabla V)^2 - \nabla^2 V\right)$.

Once $ {\bar H} $ is in form (\ref{f2}), it can be supersymmetrized following the 
prescription
\eqb 
Q_i &\to& S \,\,= Q_i \psi_i,  \nonumber
\\
Q^\dagger_i &\to& S^\dagger = Q^\dagger_i \psi^\dagger_i,   
\label{f4}
\eqf
with $\psi_i$ and $\psi_j^\dag$, elements of a Clifford algebra. While for the one dimensional 
case these elements  satisfy  $\{\psi,\psi\}=0=\{\psi^\dagger, \psi^\dagger\}$, $\{\psi,\psi^\dagger\}=1$,  
in higher dimensions  this  structure may be  more complicated. Indeed, in  three dimensions
\begin{equation}
\{\psi_i,\psi_j\} =0=\{\psi^\dagger_i, \psi^\dagger_j\},\quad
\{\psi_i, \psi^\dagger_j\}  =  \delta_{ij} +  i \kappa_{ij}
\label{f5}
\end{equation}
where $\kappa_{ij}=-\kappa_{ji}$ and $\kappa^\dagger =\kappa$.  In the present case, 
the $\{\psi_i\}_{i=1,2,3}$ defined as
\bb
\psi_i= \sigma_i \otimes \sigma_+,
\quad
\psi^\dagger_i =\sigma_i \otimes \sigma_-, \label{h1}
\ee
where ${\sigma_i}$ are the Pauli matrices and $\sigma_{\pm} =\frac12(\sigma_1\pm\imath\sigma_2)$,
furnish a realization of the previous algebra with 
\bb
\kappa_{ij} = \epsilon_{ijk}\sigma_k \otimes \sigma_3. \label{ff5}
\ee

For $\kappa_{ij}= 0$, the spinors $\psi_i, \psi_j^\dagger$ obey a standard Clifford algebra, but if $\kappa_{ij} \neq 0$, the grading of the algebra is  more involved  \cite{balantekin,kost}. For harmonic-like potentials and their variations with centrifugal potentials this is a well-established fact. The problem is more intricate for general central potentials, as was discussed in   \cite{ritte}.

The prescription in   (\ref{f4}), with $\kappa_{ij}$  given in (\ref{ff5}), generates the superalgebra
\eqb
\{S,S^\dagger\} &=& 2 {\bar H}_s, \label{g1}
\\
\left[ S, {\bar H}_s\right] &=& 0=\left[S^\dagger, {\bar H}_s\right], \label{g22}
\\
\{S^\dagger, S^\dagger\} &=&0=\{S,S\}. \label{g3}
\eqf
where \footnote{We will suppress the symbol $\,\bar{ }\, $ in the rest of the paper and it is understood that the 
Hamiltonians under discussion are shifted so that the energy of the ground state is zero.}
\begin{equation}
{\bar H}_s =
%&=& \left[\frac{1}{2}Q^\dagger_i Q_i - 2(\nabla V\times {\bf p}). \,\textbf{s} \right] \otimes \sigma_- +  \left[\frac{1}{2} Q_i %Q^\dagger_i +2 (\nabla V\times {\bf p}). \,{\bf s}\right] \otimes 
%\sigma_+ \nonumber
%\\
%&=& 
H_{-}\otimes \sigma_- + H_{+} \otimes \sigma_+, \label{g4}
\end{equation}
with
\eqb
H_{\pm} &=& -\frac{1}{2}\nabla^2 + u_\pm(\textbf{x}) \pm 2(\nabla V \times {\bf p}) \cdot\,\textbf{s}, 
\label{g5}
%\\
%H_{+} &=& -\frac{1}{2}\nabla^2 + u_+(\textbf{x}) +2(\nabla V \times {\bf p})\cdot\,\textbf{s}, 
%\label{g6}
\\
u_{\pm}(\textbf{x}) &=& \frac{1}{2}[(\nabla V)^2 {\pm} \nabla^2 V]. \label{p1}
\eqf
where  ${\bf s} =\frac12{\boldsymbol \sigma}$.
 The appearance of the spin-orbit coupling term ensures  the invariance under $T$ and introduces a magnetic field-like 
 term. To see this,  we will proceed as follows; let us  take,  for example, the component $H_{-}$ in (\ref{g5})  and 
 write it in the following suggestive form
 \eqb
 H_{-} 
 %&=& -\frac{1}{2}\nabla^2 + u_-(\textbf{x}) -2(\nabla V \times {\bf p})\cdot\textbf{s}
  %\nonumber 
 %\\
 &=& \frac{1}{2}\left[{\bf p} -2\,{\bf s}\times \nabla V \right]^2 -2\left( {\bf s}\times \nabla V\right)^2 + u_-(\textbf{x}), \label{p111}
 \eqf 
 where we have used $(\nabla V\times {\bf p})\cdot{\bf s} =({\bf s}\times\nabla V)\cdot{\bf p}$, and $
 [{\bf s}\times\nabla V,{\bf p}] =0$.  Since  $\left( {\bf s}\times \nabla V\right)^2=\frac{1}{2}(\nabla V)^2$, 
 we rewrite (\ref{p111}) as follows
\begin{equation}
H_{-} = \frac{1}{2}\left[{\bf p} -2\,{\bf s}\times \nabla V \right]^2 + \bar{u}_-({\bf x}),
\end{equation}
with
\bb
 {\bar u}_{-}({\bf x})=  -\frac{1}{2} \nabla^2 V.
 \label{pp2}
 \ee 
By  changing $V \to -V$ (and then $\bar{u}_-\to \bar{u}_+$), we obtain $H_{+}$.

To summarize,  it is possible to write $H_{\pm }$ defined in (\ref{g5})  as 
\begin{equation}
\label{hmag}
H_{\pm } = \frac{1}{2} \left[ {\bf p} \pm {\bf {\cal A}}\right]^2 +\bar{u}_{\pm}({\bf x}),
\end{equation}
with ${\bf {\cal A}} = 2{\bf s}\times \nabla V$.

 We emphasize that the term ${\bf {\cal A}}= 2{\bf s}\times \nabla V$  turns out to be a $SU(2)$ connection. Indeed,
 let us define ${\cal A}^a_i={\epsilon_i}^{a k}\partial_kV$, and therefore
 \footnote{For this part of the discussion we have used a modified notation where the group index is denoted by latin
index ${a,b,\dots}$.} ${\bf {\cal A}}_i ={{\cal A}^a}_i\,\sigma^a$ is an element of the $SU(2)$ algebra.  In order ${\bf {\cal A}}$ to be a connection, it must change under the $SU(2)$ local transformation  $U=e^{\imath \theta^a\sigma^a}$ as
\begin{equation}
{\bf {\cal A}}_i\to{\bf {\cal A}'}_i=U\,{\bf{\cal A}}_i\,U^{\dag}+\imath\,\partial_iU\,U^{\dag}.
\end{equation}
with $U^{\dag} =U^{-1}$.

Global gauge transformations are particularly interesting for our model. In fact, for $\theta^a$ constant, the  gauge transformation reads
$$
{\bf {\cal A}}\to{\bf {\cal A}'}=U\,{\bf{\cal A}} \,U^{\dag},
$$
and then, we perform the unitary transformation ${\bm {U}} =U \otimes \openone$ in $\bar{H}_s$ defined in  (\ref{g4}) as 
follows
\begin{eqnarray}
\bar{H}'_s &=&{\bm U}\,\bar{H}_s\,{\bm U}^\dag,
\nonumber
\\
&=&U\,H_{+}\,U^{\dag}\otimes\sigma_+\,+\,U\,H_{-}U^\dag\otimes\sigma_-,
\nonumber
\\
&\equiv& H'_{+}\otimes \sigma_+  \,+\,  H'_{-}\otimes\sigma_-.
\end{eqnarray}
 
Since $U({\bf p} -{\bf {\cal A}})^2\,U^\dag = ({\bf p} - U{\bf {\cal A}}U^\dag)^2$, the unitary transformed Hamiltonians
are
\begin{equation}
H'_{\pm } =  \frac{1}{2} \left[ {\bf p} \pm {\bf {\cal A}}'\right]^2 +\bar{u}_{\pm}({\bf x}).
\end{equation}

For example, consider de gauge field ${ {\cal A}}_i=A_i\, n^a \sigma^a =A_i\,{\bf n}\cdot{\boldsymbol \sigma}$, with 
${\bf n}$ a fixed direction in the $SU(2)$ algebra. The transformation

% For example $\frac{d{\bf L}}{dt}$ is 
% \[
%\frac{d{\bf L}}{dt} = \omega\, {\bf s} \times {\bf L}
%\]
%which is just the Thomas precession. It is remarkable that the interpretation of the spin-orbit coupling term as {\it magnetic field} is only fulfilled by choosing ${\bf L}= L_3 ~{ {\hat e}_3}$  for angular momentum. 

%The idea explained above is intuitive and in order to realize this argument we proceed as follows; let us consider the case of a central field and we write ${\bf L}.{\bf s}$ 
%as $$\frac{|L|}{2} {\hat {\bf n}}.\sigmavec$$
%where ${\hat {\bf n}}^2=1$ and $\ell=|L|=\sqrt{{\bf L}}$ is a constant. 

%Doing a unitary transformation  $H'=UHU^\dagger$ with 

\begin{align}
U &= U_3[\varphi/2]\,U_2[\theta/2]U_3[-\varphi/2],
\nonumber
\\
&=e^{\imath \frac{\varphi}{2}\sigma_3}\,e^{-\imath \frac{\theta}{2}\sigma_2}\,e^{-\imath \frac{\varphi}{2}\sigma_3}
\nonumber
\\
&=\left( \begin{array}{lr} 
\cos \frac{\theta}{2} & -e^{i\varphi} \,\sin \frac{\theta}{2} 
\\
e^{-i\varphi}\,\sin \frac{\theta}{2}  & \cos \frac{\theta}{2}
\end{array} \right),
\nonumber
\end{align}
and the identity (Hopf map) \cite{aitchi}
\[
U \,{\hat {\bf n}}\cdot\sigmavec\,U^\dagger = \sigma_3, 
\]
allow to transform equation (\ref{g4}) to
\bb
{\bar H}_s'= {\bar H}'_{-}\otimes \sigma_- + {\bar H}'_{+} \otimes \sigma_+, \label{g41}
\ee 
where 
\bb
{\bar H}'_{\pm } =[ -\frac{1}{2}\nabla^2 + {u}_\pm(\textbf{x})\pm (\nabla V\times {\bf p})_3\,\sigma_3]. 
\label{g511}
\ee

For the case of a central potential $V({\bf x})=V(r)$, one has 
$$ \nabla V\times{\bf p} = \frac{dV}{dr}\hat{r}\times{\bf p}
=\frac{1}{r}\frac{dV}{dr}(r\hat{r}\times{\bf p})=\frac{1}{r}\frac{dV}{dr}{\bf L},
$$
with ${\bf L}$, the angular momentum.

Therefore, in the present case, after the gauge transformations, we get for the central potential
\bb
{\bar H}'_{\pm} =[ -\frac{1}{2}\nabla^2 + {u}_\pm(\textbf{x}) \pm \frac{L_3}{r} \frac{dV}{dr}  \sigma_3], 
\label{g5111}
\ee
and since $[L_3,H'_{\pm}]=0$,  operator  $L_3$  can be replaced by its eigenvalues.
%\to \ell$  
%Finally,
%\bb
%{\bar H}'_{\pm V} =[ -\frac{1}{2}\nabla^2 + {\bar u}_\pm(\textbf{x}) - \frac{\ell}{r} \frac{dV}{dr}  \sigma_3]\otimes \sigma_{\pm}, \label{g51}
%\ee

On the other hand, in this approach, we can define the covariant derivative and the curvature in the standard way, and 
since ${\cal A}$ is a non-abelian connection, the curvature ${\cal F}_{\mu \nu}$ is not gauge invariant. However, for a topological insulator, the invariance under $T$ allows to invert the momentum and the spin simultaneously and, therefore, if we choose the spin on the axis-$3$, the two components of the constant chromomagnetic field \cite{weis} can be interpreted as a real magnetic field in the up and down directions.

The curvature can be calculated from the commutator $\left[D_i({\cal A}),D_j({\cal A})\right]$, 
\bb
\left[D_i({\cal A}),D_j({\cal A})\right] = \imath {\cal F}_{ij}, 
\ee
with $ D_i({\cal A}) = \partial_i + 2i ({\bf s}\times \nabla V)_i$ (and a similar definition for $H'_{+}$, by changing $V\to -V$)
 \[
 {\cal F}_{ij} = 2\imath\left(\epsilon_{jkl} \partial_i \partial_l V -\epsilon_{ikl} \partial_j \partial_l  V \right) s_k.
 \]

Therefore, using the arguments explained above, for the connection along $\sigma_3$, one finds
\bb
{\cal F}_{12}=\mathbb{B}= \nabla^2 V\, \sigma_3, \label{spin}
\ee
where $\nabla^2$ is the two-dimensional Laplacian and the eigenvalues of ${\cal F}_{12}$ contains the up and down components for the \lq \lq magnetic field\rq \rq \,stated above.

From these results we see that  Hamiltonian 
\bb
{\bar H}_s=\left[\frac{1}{2} ({\bf p} -{\cal A})^2 + {\bar u}_-({\bf x}) \right] \otimes \sigma_- + \left[\frac{1}{2} ({\bf p} +{\cal A})^2 + {\bar u}_+({\bf x}) \right] \otimes \sigma_+, \label{hh2}
\ee 
describes a supersymmetric system coupled to a $SU (2)$ gauge field \cite{wong,bal} and the role played by this field is to implement the topological magnetic field. Note that in three dimensions --as in anyons where one adds a Chern-Simons term in $2+1$ dimensions \cite{rich}-- we should add a Pontryagin term, but this is a total derivative in the Lagrangian and does not contribute to the Hamiltonian. However, the total derivative leaves a non-trivial topological effect;  it is responsible for the magnetoelectric effect.

Note also that,    if   $\nabla V \times {\bf p}$ varies adiabatically, then we can approximate 
\bb
{\bar H}_{\pm } = \frac{1}{2}{\bf p}^2 + {u}_{\pm}({\bf x}) \pm 2 (\nabla V \times {\bf p})\cdot{\bf s}, \label{gp1}
\ee
to
\eqb
{H}_{\pm } &\simeq &- 2(\nabla V \times {\bf p})\cdot{\bf s} \nonumber
\\
&\simeq& - {\bf B}\cdot{\bf s}, \label{gp2}
\eqf
where ${\bf B}=2\nabla V \times {\bf p}$ is interpreted as a  fictitious magnetic field. What does this approximate physically mean?

Let us assume that over  some typical length scale ${\it L}$ the following relation holds $ |\partial_i V|\gg L\,\partial_i\partial_j V$, (in one dimension this means $V'\gg LV''$)
so that the potential is of the order $V\approx V_0 +L|\nabla V|$, with $V_0$ the mean value of the potential inside the region of size $L$. In this case, the potential $u_\pm\sim (\nabla V)^2$.  For heavy modes, that is $p^2\ll m$, the condition 
$|\nabla V|\ll |{\bf p}|$, assures that the term $|\nabla V\times {\bf p}|$ dominates the complete Hamiltonian, giving rise
to the adiabatic limit previously discussed.

Using the de Broglie relation we can write an energy scale $E_{s}\sim \frac{1}{\lambda}$, characterizing the slow modes 
(highly massive modes) and also the energy scale $E_{f}\sim \frac{1}{L}$, characterizing the energy necessary to explore 
the region of size $L$. The condition $|\nabla V|\ll |{\bf p}|$, reads now 
\bb
\frac{E_f}{E_s} \ll1.
\ee

The adiabatic approximation and its topological implications for this kind of problems
was studied by Berry \cite{berry} who discovered that in the parameter 
space ${\bf B }$ the gauge potential of this system is a Dirac monopole described by 
\bb
{\cal A}_i = \frac{1}{2}\frac{B_i}{|{\bf B}|^3}. \label{gp3}
\ee
where the Berry phase $\gamma_{\pm}(C)$ satisfies
\bb 
e^{i \gamma_{\pm} (C)}= e^{\oint_C dB_i {\cal A}_i} = e^{\mp i {\frac{n}{2} \Omega (C)}},
\ee
where $\Omega$ is the solid angle subtended in the parameters space.
\section{Examples Continuous  Topological insulators  Quantum mechanics }

In order to illustrate the ideas set out above, we will discuss two examples of topological insulators that can be studied directly using our arguments.
\subsection{Harmonic oscillator in three dimensions} 
This example is a  variation of the problem studied  in \cite{zhang}. Consider
 the three-dimensional harmonic oscillator with frequency $\omega$ and mass $m$. The ground state (up to 
 a normalization constant)  is
\bb
\psi_0 =  e^{-\frac{m \omega}{2} r^2},
\ee 
with $r$ the radial coordinate. The function $V$ in  (\ref{f0}) reads  $V= \frac{m \omega}{2} r^2$, and 
the potential in  (\ref{p1}) is 
\bb
u_\pm = \frac{1}{2}\left( m^2 \omega^2 r^2\mp 3m \omega \right). 
\label{pot1}
\ee
For the supersymmetric  Hamiltonian ${\bar H}'_s$ in (\ref{g41}) (restoring the mass $m$), the ${\bar H}'_\pm$ components
(see  (\ref{g5111})) read
\begin{equation}
{\bar H}'_\pm =-\frac1{2m}\nabla^2+\frac{1}{2}\left( m \omega^2 r^2\mp 3 \omega \right)\pm\,\omega\,L_3\sigma_3,
\end{equation}
and  the Schr\"odinger-Pauli  equation  turn out to be
 \begin{equation}
 {\bar H}'_\pm\psi_\pm=E\,\psi_\pm.
 \end{equation}
Let us consider the diagonalization of ${\bar H}'_-$, that is
$$
 {\bar H}'_-\psi_-=E\,\psi_-.
$$

 We look for solutions for the spinor $\psi_-$ with the form
 \begin{equation}
\label{soloscp}
\psi_-(r,\theta,\varphi) =
\left(
\begin{array}{c}
R_-^{(+)}(r) Y_\ell^{n^{(+)}}(\theta,\varphi)
\\
R_-^{(-)}(r) Y_\ell^{n^{(-)}} (\theta,\varphi)
\end{array}
\right),
\end{equation} 
 where $Y_\ell^n(\theta,\varphi)$ 
  are the spherical harmonics. 
  
  The equation for $R_{-}^{(\pm)}(r)$ turn out to be
  \begin{equation}
  \label{oscp}
  {R_{-}^{(\pm)}}''+\frac{2}{r}{R_-^{(\pm)}}' 
  -\left[\frac{\ell(\ell+1)}{r^2} + m\omega\left(m\omega\,r^2+3\right) -2mE\mp2m\,n^{(\pm)}\omega\right]{R_-^{(\pm)}}=0.
  \end{equation}
In terms of a dimensionless variable $x=\sqrt{m\omega}\,r$, and defining $\epsilon^2 =\frac{2E}{\omega}$, 
previous equation reads
\begin{equation}
\label{osc2}
x^2\,{R_-^{(\pm)}}''+2x\,{R_-^{(\pm)}}'-\left[x^4+(3-\epsilon^2\mp2n^{(\pm)})x^2+\ell(\ell+1)\right]{R_-^{(\pm)}}=0,
\end{equation}
whose normalizable solutions  are
\bb
{R_-^{(\pm)}}(x) = x^\ell\,{e^{-\frac{x^2}{2}}}\,L^{\ell+1}_{\kappa^\pm}(x),
\ee
with $2\kappa^{(\pm)} = 3+\frac{E}{\omega}-\ell\pm n^{(\pm)}$.  The solution is square integrable for $\kappa^{(\pm)}\in \mathbb{Z}^+$. Finally, $\psi_-$ then reads (up to a normalization constant)
\begin{equation}
\label{soloscp}
\psi_-(r,\theta,\varphi) = r^\ell\,e^{-\frac{m\omega r^2}{2}}
\left(
\begin{array}{c}
L^{\ell+1}_{\kappa^{(+)}}(x)\,Y_\ell^{n^{(+)}}
\\
L^{\ell+1}_{\kappa^{(-)}}(x)\,Y_\ell^{n^{(-)}}
\end{array}
\right).
\end{equation}

Note that the energy levels  $E=\omega(\ell-3+2\kappa^{(\pm)}\mp n^{(\pm)})$ are well defined for $2\kappa^{(+)}
-n^{(+)}=2\kappa^{(-)}+n^{(-)}$ only. Condition $E>0$ imposes also $2\kappa^{(\pm)}>3-\ell\pm n^{(\pm)}$.

Consider the case for which $\kappa^{(+)}>\kappa^{(-)}$. Then, the previous restrictions imply $n^{(+)}+n^{(-)}>0$.
For the case $\kappa^{(+)}<\kappa^{(-)}$, restriction reads $n^{(+)}+n^{(-)}<0$. That is, the $L_3$ projections of
the spinor's components are not independent.

The situation for  
$\kappa^{(+)} =\kappa^{(-)}$, is similar and projections must satisfy $n^{(-)} = - n^{(+)}$. We can write the solution 
for this case as follows
\begin{equation}
\label{soloscpequ}
\psi_-(r,\theta,\varphi) = r^\ell\,{e^{-\frac{m\omega r^2}{2}}}
\left(
\begin{array}{c}
L^{\ell+1}_{\kappa}(x)\,Y_\ell^{n}
\\
L^{\ell+1}_{\kappa}(x)\,Y_\ell^{-n}
\end{array}
\right).
\end{equation}

Clearly, in all three cases, the energy can be written as $E_{\kappa^{(+)},\ell,n^{(+)}}=\omega(2\kappa^{(+)}+\ell-3-n^{(+)})=\omega(2\kappa^{(-)}+\ell-3-n^{(-)})$, and the system is degenerated. It is worthwhile to notice that in the  
absence of the term proportional to $L_3$, the solution is $n^{(\pm)}$-independent and therefore $\kappa^{(+)}=
\kappa^{(-)}$ so that the solution depends only on integers $\kappa$ and $\ell$.

The diagonalization of ${\bar H}'_+$ proceeds in a similar way. We look for a solution for the  spinor $\psi_+$ 
with the form 
\begin{equation}
\label{soloscm}
\psi_+(r,\theta,\varphi) =
\left(
\begin{array}{c}
R_+^{(+)}(r) Y_\ell^{n^{(+)}}(\theta,\varphi)
\\
R_+^{(-)}(r) Y_\ell^{n^{(-)}} (\theta,\varphi)
\end{array}
\right).
\end{equation} 
Radial components are solutions of 
  \begin{equation}
  \label{oscm}
  {R_+^{(\pm)}}''+\frac{2}{r}{R_+^{(\pm)}}' 
  -\left[\frac{\ell(\ell+1)}{r^2} + m\omega\left(m\omega\,r^2-3\right) -2mE \pm 2m\,n^{(\pm)}\omega\right]{R_+^{(\pm)}}=0,
  \end{equation}
or, in terms of dimensionless variables
\begin{equation}
\label{osc2}
x^2\,{R_+^{(\pm)}}''+2x\,{R_+^{(\pm)}}'-\left[x^4+(-3-\epsilon^2 \pm 2 n^{(\pm)})x^2+\ell(\ell+1)\right]{R_+^{(\pm)}}=0.
\end{equation}
Solutions of this equations are those in (\ref{soloscp}) while the energy turn out to be $E=\omega(2\kappa^{(\pm)}\pm n^{(\pm)} +\ell)$, from which we obtain the condition  $2\kappa^{(+)}
+ n^{(+)}=2\kappa^{(-)}-n^{(-)}$, while  $E>0$ imposes also $2\kappa^{(\pm)}+\ell>\mp n^{(\pm)}$.

 \subsection{ The Aharonov-Bohm Topological Insulator}
  
 A second example is the case of the Aharonov-Bohm effect, which we can study directly through the identification of 
 $V$ according to  
 \bb
 \nabla^2 V = \alpha\, \delta ({\bf x}_\perp), 
  \ee
 where $\alpha$ is a dimensionless coefficient defined in terms of the magnetic flux as  $\Phi_B=2\pi\alpha$ and 
 ${\bf x}_\perp$ are coordinates in two dimensions. 
 
In order to calculate $u_\pm$ we need to solve
\bb 
\nabla^2 V=  \frac{\Phi_B}{2\pi} \delta ({\bf x}_\perp),
\label{mono1}
\ee
what can be achieved, for example, by using the Green function $G(|{\bf x}-{\bf x}'|)$. Indeed,
\eqb
V({\bf x}_\perp) &=& \frac{\Phi_B}{2\pi} \int  \, G(|{\bf x}_\perp-{\bf x}'_\perp|)\delta^{(2)} ({\bf x}')\,d^2x'
 \nonumber
\\
&=&\frac{\Phi_B}{2\pi} \,G(|{\bf x}_\perp|), 
\label{mono2}
\eqf
with 
\bb
G(|{\bf x}_\perp|) = \int \frac{d^2 p}{(2\pi)^2} \frac{e^{i {\bf p}\cdot{\bf x}_\perp}}{{\bf p}^2} = \ln |\mu  {\bf x}_\perp|, \label{mono3}
\ee
with $\mu$ an energy scale.

Using (\ref{p1}) and (\ref{mono2}) the potential  is
\bb
u_\pm ({\bf x}_\perp) = \frac{1}{2} \left[ \frac{\Phi_B^2}{ 4 \pi^2|{\bf x}_\perp|^2} \pm  \frac{\Phi_B}{2\pi} \delta ({\bf x}_\perp)\right], \label{mono4}
\ee
and the Hamiltonian ${\bar H}'_\pm$ in  (\ref{g5111}) (or ( \ref{g5}))  becomes 
\bb
{\bar H}'_\pm = \frac{1}{2} \left[ -\nabla^2 + \frac{\Phi_B^2}{4 \pi^2|{\bf x}_\perp|^2} \pm \frac{\Phi_B}{2\pi}~\delta ({\bf x}_\perp) \right] \pm \frac{\Phi_B}{2 \pi |{\bf x}_\perp|^2} \sigma_3 L_3. \label{last}
\ee

The diagonalization of the supersymmetric Hamiltonian implies, as before, the following equations
\begin{equation}
{\bar H}'_\pm\,\psi_\pm=E\psi_\pm.
\end{equation}
Explicitly, in polar coordinates, this equation reads
\begin{equation}
{\bar H}'_\pm =\frac{1}{2} \left[-\nabla_\perp^2 - \partial_z^2 
+ \frac{\Phi_B^2}{4 \pi^2{\rho^2}}
 \pm 
 \frac{\Phi_B}{2\pi}\frac{\delta (\rho)}{\rho}\right]   \pm \frac{\Phi_B}{2 \pi \rho^2} \sigma_3 L_3.
\label{last1}
\end{equation}
with $\nabla_\perp^2=\rho^{-1}\partial_\rho[\rho\partial_\rho\,]+\rho^{-2}\partial_\theta^2$.

Let us consider first the ${\bar H}'_-$. For such a case we look for solutions with the form
\begin{equation}
\psi_-(\rho,\theta,z) =e^{\imath k z} \sum_{n=-\infty}^{n=\infty}\left(
\begin{array}{c}
R_-^{(+)}\,e^{\imath \theta n^{(+)}}
\\
R_-^{(-)}\,e^{\imath \theta n^{(-)} }
\end{array}
\right),
\end{equation}
where the  integers $n^{(\pm)}$ are the winding numbers.   

The radial part,  satisfies
\begin{equation}
{R_-^{(\pm)}}''+\frac1\rho{{R_-^{(\pm)}}'}-(k^2-\epsilon^2){R_-^{(\pm)}}-
\frac1{\rho^2}\left(n^{(\pm)} \mp \frac{\Phi_B	}{2\pi}\right)^2{R_-^{(\pm)}}=
 \frac{\Phi_B	}{2\pi}\frac{\delta(\rho)}{\rho}{R_-^{(\pm)}},
 \label{last1}
\end{equation}
with $\epsilon^2=2E$. For the Hamiltonian ${\bar H}'_+$, the radial part equation can be obtained from this one
by changing $\Phi_B\to-\Phi_B$.

The factor $\frac{\Phi_B}{2\pi}$ in (\ref{last1})   is the magnetic field inside the solenoid, while  the change of  signs 
in front the $\delta$-function when passing from ${\bar H}'_-$ to  ${\bar H}'_+$, shows that the system is a topological 
invariant  \cite{moo1,ab2}. The complete solution of the 
problem is the superposition of different $n^{(\pm)}$, that is, different  winding numbers for spin components up and down.
\section{Continuous Topological Insulators in High Energy Physics}

In this section, we will discuss  the motivation for introducing topological insulator techniques in high-energy physics through two
examples illustrating this possible connection.

{{We would like to draw attention to two observations that make this study worthwhile. The first one is purely theoretical}} because it is very suggestive that the quantum Hall effect and topological insulators --  well-known examples of topological phases --  are understood from a similar mathematical framework. The second reason is that the analysis of non-perturbative physics from the Born-Oppenheimer approximation viewpoint and its variants is also a similar problem. {{ At}} least from the mathematical point of view, an analysis  in this direction is fully justified.

\subsection{Dirac Oscillator Extension}
The first example we will consider is the Dirac oscillator extension discussed in \cite{moshinsky} where instead of the replacement $p_i \to\pi_i= p_i + i m \beta x_i$ for the Dirac equation, we will consider
\bb
\pi_i = \left\{
\begin{array}{lll}  \left(
p_i +  B \epsilon_{ij} x_j\otimes \sigma_3\right), &   &\{i,j\}=1,2, \label{pi}
 \\
p_3, &    & ~~~~~~i=3.
\end{array}
\right.
\ee
Here $B$ is a constant with canonical dimension $+2$ and $\sigma_3$ is defined  in the 
color space. 

Note that, similarly to the Landau problem, it holds
\eqb
\left[x_i,\pi_j\right] &=& i \delta_{ij},
 \nonumber 
\\
\left[\pi_i,\pi_j\right] &=&i {\cal F}_{ij}, \label{ss}
\eqf
and zero in all other cases and 
\bb
{\cal F}_{ij} = 2 B \epsilon_{ij} \sigma_3, \label{cur1}
\ee
is a constant curvature {{defined }} in $SU(2)$. This curvature can also be viewed as {{the one} of a diagonal constant chromagnetic field as in \cite{weisberger}.

From the Dirac Hamiltonian
\bb
H = \sigmavec\cdot\pivec  + \beta m, \label{di1}
\ee
we find that the Dirac equation for the choice of $\pivec$ becomes
\begin{eqnarray}
\label{osc33}
\bigg[ \boldsymbol{\sigma}\cdot{\bf p}\otimes\openone
+B(\boldsymbol{\sigma}\times{\bf x})\cdot\hat{z}\otimes\sigma_3\bigg]\Phi_2 
&=&(E-m)\openone\otimes \openone\, \Phi_1  ,
\\
\label{osc34}
\bigg[ \boldsymbol{\sigma}\cdot{\bf p}\otimes\openone
+B(\boldsymbol{\sigma}\times{\bf x})\cdot\hat{z}\otimes\sigma_3\bigg]\Phi_1
&=& (E+m)\openone\otimes \openone\, \Phi_2.
\end{eqnarray}

This equation can be decoupled by using standard methods and we get and we find for one of the components
\begin{equation}
\label{osc35}
\bigg[\big(
{\bf p}_\perp^2+p_3^2 + B^2{\bf x}_\perp^2
\big)\otimes\openone -2B(L_3+\sigma_3)\otimes\sigma_3\bigg]\Phi_1
=
(E^2-m^2)\openone\otimes\openone\,\Phi_1,
\end{equation}
where ${\bf p}_\perp=p_1^2 +p_2^2$.

The spectrum is exact \cite{wedirac} and can be written as 
\bb
E^2_\pm =   m^2 + p_z^2 + 2B \left(2 n_\pm + |\ell |-\ell+1\pm 1 \right). 
\label{spectrum}
\ee

We should note that in (\ref{osc33}) and (\ref{osc34}),  each operator is an array of $4\times 4$, and thus this problem corresponds to two {{copies}} of Landau ones. 
This  {{fact is reflected on}} the spectrum (\ref{spectrum}), which  {{on the other hand,}}  apparently violates particle-antiparticle symmetry. However, {{the condition $E_-=E_+$ implies $n_-  - n_+ = 1$, and  the spectrum  becomes
symmetric under  $n_- \leftrightarrow n_+$ and 
$\ell \leftrightarrow -\ell$, making the particle-antiparticle asymmetry only apparent.}}

\subsection{Magnetoelectric Effect and Dark Matter}

In this section, we outline the implications that continuous topological insulators might have for dark matter (DM) physics. Consider a dark matter clump modeled as a continuous dielectric material of volume $V$ located in a region of 
space. Observations indicate that DM interacts with ordinary matter through gravity, while other interactions are 
negligible. Indeed, the search for a model for DM with such properties is one of the main research topics in physics nowadays.

On the other hand, dark matter  can be subject to external electromagnetic radiation, and even considering
that the interaction can be safely neglected, the possibility of topological effects can not be discarded from the beginning.
Due to this effect,  the  DM  might behave as a dielectric which, in principle,   could be polarized and magnetized. Let us
investigate how could this happen.

The total action (radiation + dielectric) is  given by 
\bb
S = \int d^4 x\left( \frac{\epsilon}{2} {\bf E}^2 - \frac{1}{2 \mu} {\bf B}^2 + \cdots \right), \label{den1}
\ee
where $\epsilon$ and $\mu$ are the dielectric and magnetic constant and $\cdots$ denotes  higher-order power in ${\bf E}$ and ${\bf B}$, and terms of the form $({\bf E}\cdot{\bf B})^m$ with $m\in \mathbb{Z}^+$.
From (\ref{den1})  the polarization and magnetization are 
\eqb
{\bf P}&=& \frac{\delta S}{\delta {\bf E}} =\epsilon {\bf E} + {\bf F}({\bf E},{\bf B}), \label{po1}
\\
{\bf M}&=&\frac{\delta S}{\delta {\bf B}} =-\frac{1}{\mu} {\bf B}+ {\bf G}({\bf E},{\bf B}), \label{po2}
\eqf
where ${\bf F}$ and ${\bf G}$ are functions that, for the arguments we will give, we do not need to make explicit.

Let us first notice that the energy has the form kinetic energy plus potential one, and we are interested in the "quasi-static" region such that $K +V \approx 0$. Then
\bb
S \sim \int  ({\bf E}\cdot{\bf B})^m, 
\ee
and $\bf p$ and $\bf M$ become
\bb
{\bf P} \sim m \,({\bf E}\cdot{\bf B})^{m-1} {\bf B}, ~~~~~~~~~{\bf M} \sim m \,({\bf E}\cdot{\bf B})^{m-1} {\bf E}, \label{gene1}
\ee
therefore if $m=1$, (\ref{gene1}) reduces to
\bb
{\bf P} \sim   {\bf B}, ~~~~~~~~~{\bf M} \sim {\bf E}, \label{gene2}
\ee
which is the magnetoelectric effect.

We must observe that for $m=1$,
\bb
S = \int_{\cal M} d^4x ~{\bf E}\cdot{\bf B} = \int_{\cal M} d^4 x\, F_{\mu \nu} {\tilde F}_{\mu \nu}, \label{gene3}
\ee
which is a total derivative different from zero and  is  proportional to an integer if the manifold ${\cal M}$ is topologically non-trivial (Pontryaguin theorem).

The effects of a non-trivial topology in a continuos topological insulator are explained from this term. The arguments based on supersymmetry presented in this work are efforts in this direction. 

More details of this approach are currently in progress.

\section{ Conclusions}

The analysis of topological insulators in the context of quantum mechanics (and eventually quantum field theory) is an {{exciting}} possibility that can contribute   {{to new physics ideas}} beyond  condensed matter physics. In this paper, we have proposed a route on how to proceed.

The first conclusion  from our approach is that the need to include a spin-orbit term does not depend on the energy scale, as can be seen, using supersymmetry {\it a l\'a} spin-orbit. Therefore, the connection between supersymmetry and topology is {\it an emergent} consequence, and it will certainly require a more rigorous analysis for other applications. The second conclusion is that these results open the possibility of analyzing
other quantum mechanical systems from this perspective and their implications in noncommutative quantum mechanics and integrable quantum systems seem feasible.

Finally, we consider the extensions to relativistic problems and outline the approach for both cases, which seem 
promising to apply, particularly in dark matter physics.

\noindent 

\section*{Acknowledgements}

  One of us (J.G.) thanks Prof.  Andreas Ringwald for the discussions and hospitality at DESY (Hamburg) and the Alexander von Humboldt Foundation by support. We would like also to thank H. Falomir, R. Ramirez and J. Zanelli for helpful discussions. This research was supported by DICYT 042131GR (J.G.) and 041931MF (F.M.).

\end{document}